\documentclass[%
reprint,
superscriptaddress,
 amsmath,amssymb,
 aps,
floatfix,
]{revtex4-2}

\usepackage{graphicx}
\usepackage{dcolumn}
\usepackage{bm}

\usepackage[colorlinks=true,
            linkcolor=blue,
            citecolor=blue,
            urlcolor=blue]{hyperref}

\newcommand{\mst}{MnSb\textsubscript{2}Te\textsubscript{4}}
\newcommand{\mbt}{MnBi\textsubscript{2}Te\textsubscript{4}~}
\newcommand{\Hy}{H\textsuperscript{+}~}

\begin{document}

\preprint{APS/123-QED}

\title{Hydrogen-induced modification of the local charge density in \mst}

\author{Kyriaki Samioti}
\affiliation{Université Paris-Saclay, CNRS, Laboratoire de Physique des Solides, 91405, Orsay, France}

\author{Weiyan Qi}
\affiliation{Université Paris-Saclay, CNRS, Laboratoire de Physique des Solides, 91405, Orsay, France}

\author{Stefano Ponzoni}
\affiliation{Laboratoire des Solides Irradiés, CEA/DRF/lRAMIS, CNRS, Ecole Polytechnique, Institut Polytechnique de Paris, Palaiseau, France}

\author{Antonio Salame}
\affiliation{Université Paris-Saclay, CNRS, Laboratoire de Physique des Solides, 91405, Orsay, France}

\author{Entela Buzi}
\affiliation{Department of Physics, The City College of New York - CUNY, New York
10031, USA}

\author{Afrin Nahar Tamanna}
\affiliation{Department of Physics, The City College of New York - CUNY, New York
10031, USA}

\author{Ayesha Lakra}
\affiliation{Department of Physics, The City College of New York - CUNY, New York
10031, USA}

\author{Steven Francis Udaundo}
\affiliation{Université Paris-Saclay, CNRS, Laboratoire de Physique des Solides, 91405, Orsay, France}

\author{Ding Pei}
\affiliation{Synchrotron SOLEIL, L’Orme des Merisiers, Saint-Aubin, France}

\author{Pavel Dudin}
\affiliation{Synchrotron SOLEIL, L’Orme des Merisiers, Saint-Aubin, France}

\author{Marta Zonno}
\affiliation{Synchrotron SOLEIL, L’Orme des Merisiers, Saint-Aubin, France}

\author{François Bertran}
\affiliation{Synchrotron SOLEIL, L’Orme des Merisiers, Saint-Aubin, France}

\author{Amina Taleb-Ibrahimi}
\affiliation{Synchrotron SOLEIL, L’Orme des Merisiers, Saint-Aubin, France}

\author{Lia Krusin-Elbaum}
\affiliation{Department of Physics, The City College of New York - CUNY, New York
10031, USA}
\affiliation{City University of New York Graduate Center, New York, New York 10016,
USA}

\author{Luca Perfetti}
\affiliation{Laboratoire des Solides Irradiés, CEA/DRF/lRAMIS, CNRS, Ecole Polytechnique, Institut Polytechnique de Paris, Palaiseau, France}

\author{Evangelos Papalazarou}
\affiliation{Université Paris-Saclay, CNRS, Laboratoire de Physique des Solides, 91405, Orsay, France}

\author{Marino Marsi}
\affiliation{Université Paris-Saclay, CNRS, Laboratoire de Physique des Solides, 91405, Orsay, France}

\date{\today}

\begin{abstract}

We investigate the effects of ionic hydrogen (H$^+$) intake on the occupied and unoccupied states of \mst, a van der Waals magnetic material. Using angle-resolved photoemission spectroscopy (ARPES) and time-resolved ARPES, combined with resonant and core level photoemission, we reveal that hydrogen binds near specific lattice sites, locally redistributing charge density and reshaping the valence band spectral weight. While the band gap and picosecond carrier dynamics show only minor changes upon hydrogenation, the conduction band exhibits a striking stoichiometry-dependent response. These findings indicate that hydrogen interacts strongly with the intrinsic defects in the lattice, acting as a selective dopant in \mst. Our work provides a spectroscopic basis for using hydrogen as a tool for the control of the electronic and magnetic properties in layered 2D materials.

\end{abstract}

\maketitle


\section{\label{sec:Intro}INTRODUCTION}

Intrinsic magnetic topological insulators (MTIs), with the general formula MA\textsubscript{2}B\textsubscript{4} (M = transition-metal or rare-earth, A = Bi or Sb, B = Te, Se or S), naturally combine magnetism
and topology into a van der Waals structure, enabling the emergence of exotic quantum phases. In two dimensions this manifests as Chern insulators, hosting the quantum anomalous Hall effect (QAHE), while in three dimensions it can give rise to axion insulator states \cite{Li2019rev}. Among MTIs, \mbt(MBT) has been established as an n-type antiferromagnetic (AFM) topological insulator (TI) \cite{Chen2019, Chen2019_2, Li2019, Swatek2020}, while substitution of Bi
with Sb leads to an n-to-p carrier type transition accompanied
by more diverse magnetic and electronic properties
\cite{Chen2019_2, Zhang2011} that are highly sensitive to intrinsic
defects and dopants. 

Like MBT, \mst~(MST) crystallizes in a rhombohedral layered structure
composed by Te(1)-Sb-Te(2)-Mn-Te(2)-Sb-Te(1) septuple layers separated
by van der Waals (vdW) gaps, as shown in Fig.~
\hyperref[fig:ARPES]{\ref{fig:ARPES}a}. In the a-b plane, atoms
are arranged on a triangular lattice \cite{Murakami2019,Li2022}, conventionally described using hexagonal axes with a correspondingly hexagonal surface Brillouin zone (Fig. \hyperref[fig:ARPES]{\ref{fig:ARPES}b}). The two sibling compounds share the same dominant defect type: Mn-Bi/Sb antisites. In MST antisite disorder is considerably more pronounced because of the similar electronegativities and atomic sizes of Mn and Sb \cite{Xiong2023}. In MBT, the concentration of Mn\textsubscript{Bi} and Bi\textsubscript{Mn} antisites generally remains below 5\% and 20\%, respectively, while in MST, Mn\textsubscript{Sb} and Sb\textsubscript{Mn} antisites can reach 15\% and 40\% \cite{Liu2021,Hu2024}. 

This abundance in antisites has immediate implications on the electronic and magnetic properties of MST. The presence of
Mn atoms in the Sb layers, introduce competing exchange
interactions along the c-axis, perpendicular to the atomic
layers, allowing MST to stabilize in either a ferromagnetic
(FM) or an antiferromagnetic (AFM) ground state
depending on the stoichiometry \cite{Hu2024,Yan2021,Xi2025}. In particular, Mn-rich
samples, which contain higher concentration of antisite
defects, tend to favor the FM state \cite{Liu2021,Li2022}. As for the topological character of MST, it remains in active debate, with numerous theoretical studies reporting both topological and trivial phases \cite{Eremeev2021,Murakami2019,Zhou2020, Li2021,Liu2021,Chen2020}, a discrepancy largely attributed to the material’s sensitivity to antisite defects. Defects, thus, make MST highly tunable, but also challenging to control.

Hydrogenation offers a way to exploit this sensitivity. \citet{Tamanna2024} showed that hydrogen intake and release is
a reversible, non-destructive process capable of tuning
both the magnetic and topological properties of FM
MST. \Hy incorporation decreases the majority p-type carrier
density and switches the long-range magnetic order from
FM to AFM, while subsequent release through mild
annealing tilts the Weyl nodes lying on the $k_x - k_y$ plane.
Although hydrogenation is widely practiced in semiconductor
applications \cite{Kiyoi2026,Myers1992,Pearton1987} and, more recently, on a
variety of other systems \cite{Elias2009, Shibuya2016, Wang2019, Li2023}, its effect on the band structure of a magnetic quantum material has never been directly observed.

Here, we provide the first spectroscopic study of \Hy~intake in FM and AFM MST and we show its effects on the electronic band structure via a combination of photoemission techniques. Using angle-resolved photoemission spectroscopy (ARPES) and time-resolved ARPES (trARPES), supported by core-level and resonant photoemission, we show that the incorporation of hydrogen in the crystal structure
manifests as a redistribution of spectral weight between Mn-related bands, accompanied by a subtle changes in the band gap and electron dynamics. This behavior points to hydrogen locally modifying the bonding character at charged defect sites by passivating them. Samples of different stoichiometry responded differently
to hydrogenation, an effect that we attribute to the link between the total Mn content and the concentration of Mn-Sb defects in MST. Taken together, these results establish how hydrogen reshapes the electronic structure through its interaction with the intrinsic defects in MST and point to hydrogen as a reversible tool for engineering the properties of quantum materials.

\section{\label{sec:Mathods}METHODS}

High quality \mst~ single crystals were grown by flux method \cite{Tamanna2024}. The hydrogenation step was implemented according to the protocol established in \cite{Tamanna2024} and \cite{Deng2022}. The immersion into the aqueous HCl solution was carried out for 46 and 93 hours for the FM (Mn\textsubscript{0.93}Sb\textsubscript{2.07}Te\textsubscript{4}) and the AFM (Mn\textsubscript{0.77}Sb\textsubscript{2.25}Te\textsubscript{4}) samples respectively. For all the photoemission-based experiments, i.e., ARPES, trARPES and XPS, the samples were prepared in situ by cleaving under high vacuum ($<1\times 10^{-10}$~mbar) at $300~K$.

ARPES experiments where carried out at the CASSIOPEE beamline of SOLEIL synchrotron, at low temperatures (T=15-20 K), with an overall energy resolution of 30 meV. $\mu$-XPS and $\mu$-ARPES measurements where performed the ANTARES beamline of synchrotron SOLEIL at T = 75 K. The beam was focused down to 5 $\mu$m using a capillary mirror, enabling higher photon flux and access to a broad range of photon energies \cite{Avila2023}.

The trARPES measurements where carried out at 130 K using the FemtoARPES experimental setup \cite{Faure2012}, pumping with pulses centered 1.55 eV and with 20$\mu J/cm^2$ and 400$\mu J/cm^2$ of fluence. For the case of hydrogenated samples, the lowest fluence was used in order to minimize heating effects at the surface. The fourth harmonic, centered at 6.2 eV and generated by sum frequency generation in BBO crystals ($\beta$-BaB\textsubscript{2}O\textsubscript{4}) was used as a probe either at p- or s-polarization. The temporal resolution, determined by the cross-correlation between pump and probe was $\approx100$ fs. The total energy resolution was 100 meV as determined by measuring the Fermi edge on a polycrystalline Au sample, while the angular resolution was better than 0.1$^o$ \cite{Faure2012}.

\section{\label{sec:results}RESULTS \& DISCUSSION}

\begin{figure*}
 \includegraphics{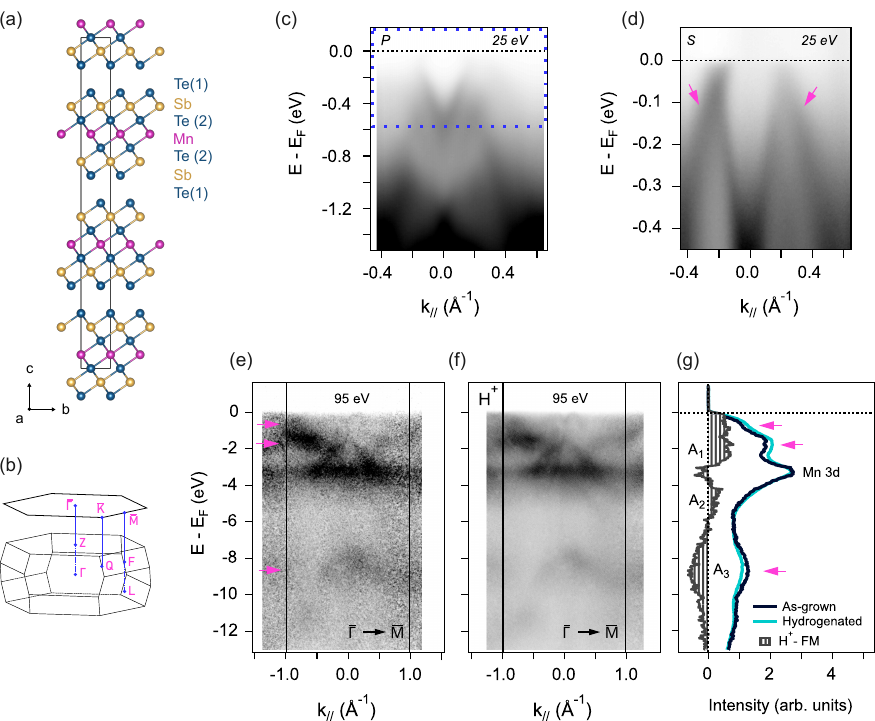}
\caption{\label{fig:ARPES} (a) Crystal structure of \mst. Te(1) and Te(2) are the two inequivalent Te sites within the septuple layer. (b) Brillouin Zone of the crystal. (c) and (d) Band dispersion measured at T = 16 K with hv=25 eV and P and S polarization respectively along $\Gamma - M$. (e) Band dispersions measured at T=77 K using 95 eV photon energy and p-polarized light for the as-grown FM MST sample and (f) the hydrogenated sample. (g) Comparison of the k-integrated intensity between the two samples. The arrows indicate the energies where a change in spectral weight is recorded.}
\end{figure*}

Figs. \hyperref[fig:ARPES]{\ref{fig:ARPES}c} and \hyperref[fig:ARPES]{d} show the low-energy band structure of FM \mst~measured along $\overline\Gamma-\overline{M}$ using p- and s-polarized light of $h\nu=25~eV$. The valence band crosses the Fermi level $E_F$, indicating strong p-type behavior and and is qualitatively similar to the one of
relative compounds in the family \mst/ (Sb\textsubscript{2}Te\textsubscript{3})\textsubscript{n}~with SL (septuple layer) terminations \cite{Huan2021, Xi2025}. A clear polarization dependence is observed. The bands indicated with arrows in Fig. \hyperref[fig:ARPES]{\ref{fig:ARPES}d} are visible only with s-polarization, whereas the V-shaped band is observed with both s- and p. This behavior originates from photoemission matrix-element effects, suggesting that both in-plane and out-of-plane orbitals contribute to the low-energy electronic structure. The $p_z$ orbitals of Te and Sb dominate the valence band and the conduction band close to $E_F$, and are responsible for the band inversion and the appearance of topological surface states in the case of TI MST \cite{Li2019rev}. In addition, even though Mn d-bands are usually deeper in conduction band, as they are highly localized, they contribute to the lower energy bands through Te-Mn-Te hybridization \cite{Xiong2023,Yan2019,Eremeev2017}.

Using higher photon energies, we probed the full Brillouin zone along the $\overline\Gamma-\overline{M}$ direction and compared the as-grown and hydrogenated FM sample. Figs. \hyperref[fig:ARPES]{\ref{fig:ARPES}e} and \hyperref[fig:ARPES]{\ref{fig:ARPES}f} show the band structures of both samples acquired with p-polarized light at 95 eV photon energy. At first glance, no striking differences are visible between the two samples. In both cases the Mn 3d flat band is clearly visible at 3.2 eV binding energy, characteristic of the highly localized Mn 3d orbitals. To better resolve the subtle differences between the two, the EDCs taken as an average over the momentum range indicated by the black lines in Figs.~\hyperref[fig:ARPES]{\ref{fig:ARPES}e,f} is shown in Fig.~\hyperref[fig:ARPES]{\ref{fig:ARPES}g}. With the intake of H$^+$, we observe a transfer of spectral weight from $E_B$=8.5 eV towards the bands next to the Fermi level and at the proximity of the Mn 3d band which appears to be slightly broader. The total sum of the areas below the integrated profiles equals zero ($A_1+A_2+A_3=0$), indicating the conservation of the total spectral weight in the specific region of the Brillouin zone. This suggests that hydrogenation modifies locally the charge density around specific lattice sites, altering the hybridization between the Mn 3d and the Te/Sb p orbitals.

Additionally, upon \Hy intake, the valance band remains intact with respect to the Fermi level $E_F$, with no evident shift in the band positions. According to \cite{Tamanna2024}, when \Hy~is diffused in the structure, it donates electrons to the system, effectively reducing its p-type character. It is therefore expected that this effective electron doping would shift the chemical potential away from the valence band \cite{Zhao2016}, manifesting as a rigid shift of the bands toward lower binding energies. Such behavior is typical of alkali metal doping. The evaporation of alkali metals on the surface, such as K or Cs, creates a two-dimensional electron gas (2DEG) at the surface that acts as a reservoir of itinerant electrons. This accumulation of charge at the surface builds up a uniform dipolar electric field, resulting in band bending and the apparent rigid shift of the bands \cite{Pinto2014, Zhang2023}. However, such behavior is not observed here. In contrast, \Hy is distributed into the bulk and thus, no macroscopic field is built up at the surface, keeping the chemical potential pinned, consistent with the absence of any observable shift of the bands in our data.

\begin{figure}
\includegraphics{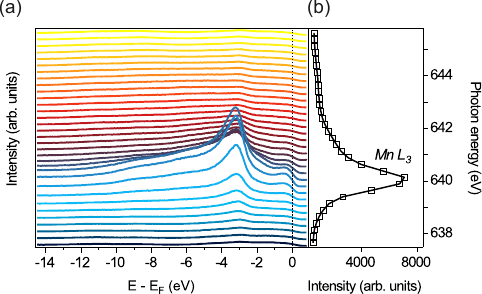} 
\caption{\label{fig:ResPhotoem} Resonant photoemission in ferromagnetic (FM) MST. The photon energy is tuned near the Mn $L_3$ absorption edge.}
\end{figure}

Identifying the orbital character of the observed bands is key to understanding how hydrogen incorporation modifies the density of states at specific energies. Direct experimental evidence on that question are obtained through resPES experiments, thanks to its enhanced element- and orbital-sensitivity. Fig. \hyperref[fig:ResPhotoem]{\ref{fig:ResPhotoem}a} shows a series of valence band spectra for FM MST obtained by resonant excitation at photon energies near the Mn $2p\rightarrow3d$ absorption edge in \hyperref[fig:ResPhotoem]{b}. These measurements reflect the partial DOS, projected on the Mn 3d orbitals, revealing two main components, namely at 0.5 eV and 3.2 eV below $E_F$, and a broad feature in the range between -6 eV and -9 eV of binding energies.

Coming back to Fig. \hyperref[fig:ARPES]{\ref{fig:ARPES}c}, it is now clear that the bands affected by hydrogen are the ones related to the Mn 3d states. Interestingly, the localized states forming the flat band at -3.2 eV binding energy are not affected by hydrogen. This points to hydrogen primarily modifying the delocalized states that participate in the hybridization with Te or Sb orbitals, forming dispersive bands, rather than atomic-like ones. To further develop this argument, hydrogen enters the lattice as a charged species H$^+$, while the lattice itself finally relaxes into a charge-neutral state. For this state to be thermodynamically stable, \Hy should not exist interstitially as a free uncompensated proton but rather create bonds with the atoms of the host lattice. Bond formation is possible only at negatively-charged sites, i.e. at acceptor-like defects. In MST such defects are primarily Mn or Sb vacancies (V\textsubscript{Mn} and V\textsubscript{Sb}) and Mn\textsubscript{Sb} antisite defects \cite{Xiong2023}. Within a more general framework, such behavior reflects hydrogen's amphoteric nature: in semiconductors \Hy can occupy different charge states during the diffusion process, passivating defects or dangling bonds by donating or accepting charge depending on the local bonding environment \cite{Kiyoi2026,Myers1992,Pearton1987}.

\subsection*{\label{sec:dynamics}Band gap and out-of-equilibrium dynamics}

\begin{figure*}
\includegraphics{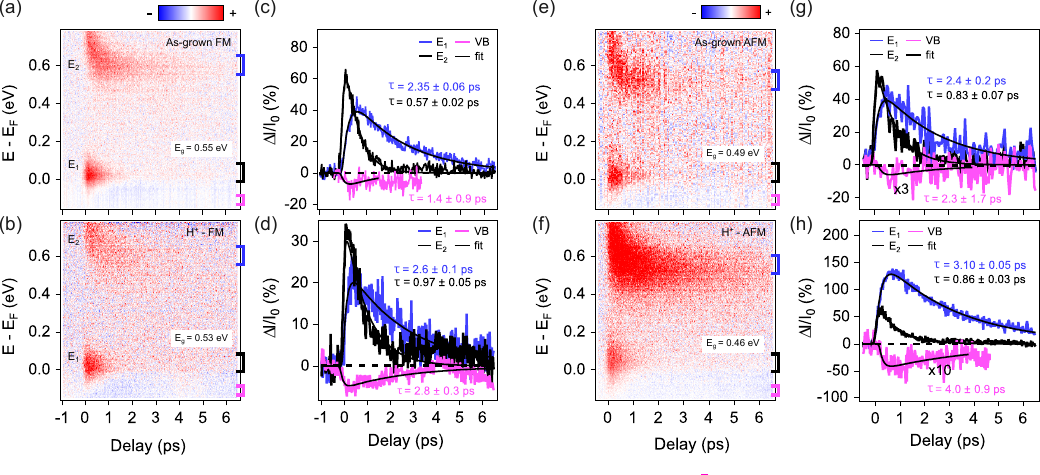}
\centering
\caption{\label{fig:dynamics} trARPES measurements taken with P-polarized probe pulses at T = 130 K. (a) $k$-integrated spectra of the as-grown MST as a function of pump-probe delay obtained by subtracting at each energy the equilibrium EDC at negative delays, $I_0$. Red and blue correspond to the gain and loss of electrons with respect to the equilibrium. The integration was done in a window $\Delta k = 0.3~\AA^{-1}$ centered at $\Gamma$. (b) Similar to (a) for the hydrogenated sample. (c) and (d) time evolution of the spectral weight obtained by integrating over 0.1 eV, as indicated by the colored brackets in (a) and (b) respectively. (e) and (f) similar to (a) and (b) for the AFM sample. (g) and (h) similar to (c) and (d) for the AFM sample.}
\end{figure*}

Having established the effects of \Hy intake on the equilibrium band structure, we investigate next the impact on the unoccupied states by means of trARPES measurements. Here, FM MST (Mn\textsubscript{0.93}Sb\textsubscript{2.07}Te\textsubscript{4}) is compared with an AFM sample, less rich in Mn (Mn\textsubscript{0.77}Sb\textsubscript{2.25}Te\textsubscript{4}). Through these experiments, the correlation between doping effects and the amount of antisite defects present in the structure are examined. Fig. \ref{fig:dynamics} shows the transient carrier population $\Delta I/I_0$ and the subsequent relaxation of the unoccupied states in the two samples before and after hydrogenation. For both samples, FM and AFM (Fig. \hyperref[fig:dynamics]{\ref{fig:dynamics}a} and \hyperref[fig:dynamics]{e}), upon the arrival of the pump pulse at zero delay, electrons are promoted into two distinct energy regions: one right above the $E_F$ (E\textsubscript{1}) where the valence band crossing was previously observed and another at higher energy, corresponding to the bulk conduction band (E\textsubscript{2}). It is noted that in region E\textsubscript{1} a finite population below $E_F$ is observed, pointing towards the presence of transient electrostatic effects acting on the surface. Similar features are observed in all samples studied here but are not considered further in the present analysis. Upon pumping, the excited electron population rapidly thermalises within $\approx$~150 fs via electron-electron intraband scattering, reaching the minimum of the conduction band (CBM) at 0.55 eV and 0.5 eV binding energy in the FM and the AFM sample, respectively. From there, phonon-assisted recombination follows on a time scale of $\approx$~2.4 ps, similar for the two cases, as shown in Figs. \hyperref[fig:dynamics]{\ref{fig:dynamics}c} and \hyperref[fig:dynamics]{g}. The recombination of the excited electrons occurs between the holes right below $E_F$, as indicated by the similar dynamics of the holes in magenta. The relatively high fitting uncertainty arises from the weak transient intensity recorded within this limited momentum range, accessible with the 6.2 eV probe ($\pm0.17~\AA^{-1}$). The faster relaxation from the E\textsubscript{1} state (in black) compared to E\textsubscript{2} is consistent with its proximity to the valence states, facilitating the recombination with the photoexcited holes close to $E_F$. The size of the band gap, defined here as the energy separation between E\textsubscript{1} and E\textsubscript{2}, is 0.554 eV for the FM and 0.487 eV for the AFM samples, respectively. These values are consistent with the work of \citet{Xi2022} where a 7.2 eV probe was used, although here the gap in FM MST is slightly smaller. The 100 meV difference in the gap size of the FM and the AFM samples reflects the difference in Mn content and, consequently, the concentration of antisite defects \cite{Liu2021}.

Regarding the effect of \Hy intake, the recombination dynamics remain unchanged in the FM sample, while they become slower by 1 ps in the AFM sample (Figs. \hyperref[fig:dynamics]{\ref{fig:dynamics}d} and \hyperref[fig:dynamics]{h}). For surface excited electrons, defects are primarily elastic scattering centers and hence they contribute to increasing the scattering rate (faster decay time). Considering that hydrogen heals intrinsic defects in MST, it is expected to decrease their total concentration and therefore intensify the presence of hot electrons at the conduction band minimum. Assuming also that the HCl dosing time is sufficient to reach saturation, the effect in AFM should be substantially larger than in FM, as it is indeed observed. Regarding the gap size, a reduce by $\approx$ 30 meV is observed upon hydrogenation in both cases, though this effect remains subtle.

\subsection*{\label{sec:DOS}Hydrogen-induced spectral weight redistribution}

\begin{figure*}
\includegraphics{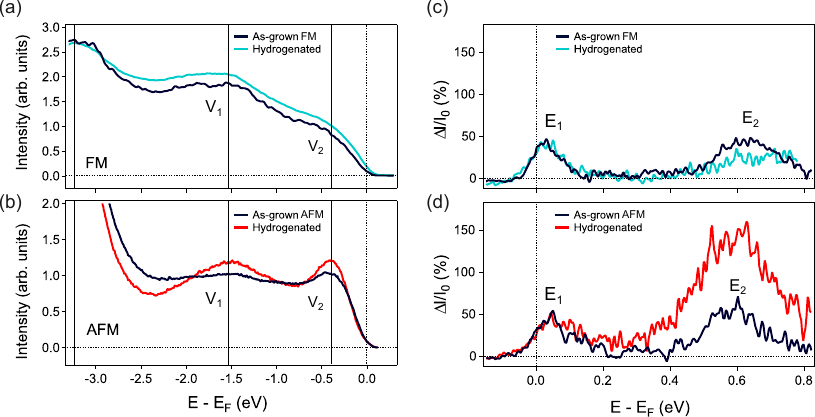}
\centering
\caption{\label{fig:FMvsAFM} (a) k-integrated photoemission yield along $\overline{M}-\overline{\Gamma}-\overline{M}$ of the FM as-grown and hydrogenated sample extracted from ($E_{kin},k_x,k_y$) maps. (b) Similar for the AFM sample. (c) k-integrated intensity of FM and hydrogenated FM. The curve was obtained at $\Delta\tau$ delay where the intensity signal of the conduction band minimum is maximum. (d) Similar for the as-grown AFM and hydrogenated samples. }
\end{figure*}

The k-integrated spectra for static and trARPES are presented in Fig. \ref{fig:FMvsAFM}. For both FM and AFM samples it is evident that hydrogenation results in the increase of spectral weight in the low energy band structure below $E_F$, particularly at the V\textsubscript{1} and V\textsubscript{2} regions as shown in Figs. \hyperref[fig:FMvsAFM]{\ref{fig:FMvsAFM}(a)} and \hyperref[fig:FMvsAFM]{(b)}. From DOS calculations it is known that the valence band maximum is dominated by the partially hybridized Te 5p with the Mn 3d states, while the lowest part of the conduction band is a mixture of Mn 3d and Sb 5p states \cite{Xiong2023,Yan2019,Eremeev2017}. As discussed above, static ARPES measurements demonstrate that during the diffusion process, \Hy interacts with negatively charged defects, resulting in the the addition of states in the low-energy valence band, irrespectively of the magnetic ground state of the sample.

    The effect on the unoccupied states, shown in  Figs. \hyperref[fig:FMvsAFM]{\ref{fig:FMvsAFM}c} and \hyperref[fig:FMvsAFM]{d}, is even more striking. The presented EDCs are extracted from the transients shown in Fig. \ref{fig:FMvsAFM}, at pump-probe delay, where the conduction band minimum intensity is maximized for each sample. In the FM sample, hydrogenation leads to the decrease of spectral weight at E\textsubscript{2} by half relative to the as-grown sample. At a striking contrast, in the AFM case the effect is reversed and more pronounced, showing three times higher gain upon hydrogenation.

Such effects in the unoccupied states indicate a change in the number of states around the conduction
band minimum, similar to what is observed in the valence band. Up to now there has been strong
evidence that \Hy modifies the Sb-Te-Mn hybridization
chain, suggesting that it preferably stabilizes near the
highly electronegative Te sites which participate in the
octahedral coordination of both Sb and Mn. In general,
\Hy is expected to locate itself around negatively
charged defects and thereby passivating them. One type
of defects, proposed in the work of \citet{Tamanna2024},
is dangling Mn-Te bonds, associated with the under-coordinated and negatively charged Mn\textsubscript{Sb} antisites \cite{Xiong2023} (Mn at an Sb site), located close to the van der Waals gap. The exposed negative charge, expected near the Te atom, makes these sites suitable for \Hy to
bind to. However, this mechanism alone, is insufficient
to explain the opposite trends between the two samples.
Since both measurements where performed in the paramagnetic
phase, where no spin-split bands are expected,
the only distinction between the two as-grown samples is
their stoichiometry, and consequently the concentration
of their intrinsic defects. At this point, we should consider that
MST has a rich landscape of intrinsic charged defects,
not limited to Mn-Sb antisites, whose formation energies
differ between Mn-rich and Mn-poor samples \cite{Xiong2023}.
Therefore, the preferential \Hy sites may depend on stoichiometry in a more complex manner, calling for a more detailed
theoretical support in order to fully understand the underlying mechanisms taking place.

\subsection*{\label{sec:XPS}Core level characterization}

\begin{figure}[htbp]
\includegraphics{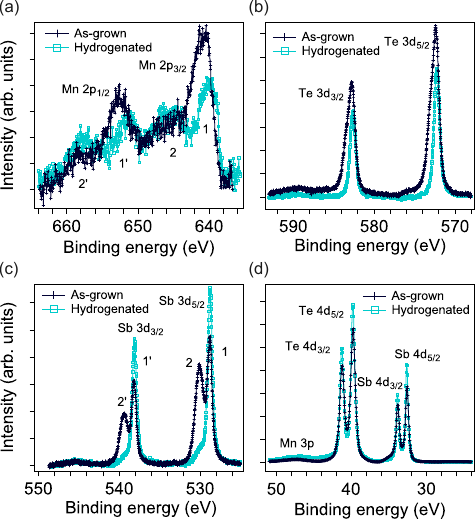} 
\caption{\label{fig:XPS} X-ray photoelectron spectroscopy (XPS) in ferromagnetic (FM) as-grown and hydrogenated MST. (a) Te $4d$ and Sb $4d$ core levels measured with 95 eV photon energy, (b) Mn $2p$, (c) Te $3d$ and (d) Sb $3d$ core levels measured with 850 eV photons. The Shirley background is subtracted from all the spectra.}
\end{figure}

XPS measurements on the as-grown FM and hydrogenated MST samples (Fig. \ref{fig:XPS}) provide direct evidence that hydrogen modifies the chemical and electronic environment around specific lattice sites, supporting our interpretation.

The Mn 2p spectra of Fig.~\hyperref[fig:XPS]{\ref{fig:XPS}a} show four main peaks: 2p\textsubscript{3/2} and 2p\textsubscript{1/2} spin-orbit components (1/1$'$) with branching ratio 1/2, and their satellite peaks (2/2$'$), originating from charge transfer between the 3d shell of Mn and the outer shell of Te during the photoemission process. Peak 1/1$'$ consists of two peaks, corresponding to Mn\textsuperscript{2+} in the Mn-Te bond and to Mn atoms with reduced-valence due to broken Mn-Te bonds \cite{Iwanowski2010,Li2022}. Upon hydrogenation, the main components shift to lower binding energy, indicating increased electron density around the Mn sites. The satellites (2/2$'$) become more pronounced without undergoing any shift, pointing to a modification of the metal (Mn) to ligand (Te)
hybridization. The Te 3d spectra of Fig.~\hyperref[fig:XPS]{\ref{fig:XPS}b} decompose into two peaks, attributed to the two inequivalent Te sites in the unit cell, namely Te atoms located adjacent to the van der Waals gap Te(1) and Te atoms neighboring the Mn layer Te(2). Upon hydrogen intake, these components shift relative to one another, leading to an apparent narrowing of the overall Te 3d feature. This indicates that hydrogen modifies the local chemical environment of the two Te sites in a site-dependent manner, consistent with a change in the Mn-Te bonding environment. This behavior is consistent with the high electronegativity of Te favoring \Hy coordination close to the Te sites \cite{Tamanna2024, Deng2022}. Additionally, the Sb 3d spectra in Fig. \hyperref[fig:XPS]{\ref{fig:XPS}c} consist of two components, labeled 1(1$'$) and 2(2$'$), separated by 1.2 eV within each spin-orbit doublet. Although the higher binding energy component resembles the spectral signature of oxidized Sb, this assignment is unlikely, since UHV cleavage conditions prevented any exposure to air. This component is therefore tentatively attributed to Sb\textsubscript{Mn} antisites. Upon hydrogenation, its intensity decreases without an appreciable energy shift, suggesting that hydrogen modifies or passivates the local environment associated with these antisites. Finally, Fig. \hyperref[fig:XPS]{\ref{fig:XPS}d} shows the 4d Te and Sb core levels at a kinetic energy corresponding to higher surface sensitivity: the comparison with the 3d peaks detected in a more bulk sensitive mode confirms that the double features in the core levels of Figs. \hyperref[fig:XPS]{\ref{fig:XPS}b} and \hyperref[fig:XPS]{c} are related to bulk effects on Te and Sb atoms.

\section{\label{sec:Conclusions}CONCLUSIONS}

To conclude, we showed that hydrogen interacts with the acceptor-like defects in MST, compensating for their excess negative charge. In particular, our results reveal that hydrogen preferably binds to defects related to Mn and Te lattice sites, therefore affecting the local Sb-Te-Mn hybridization network, as evidenced by the changes in the valence and conduction bands and the core-level spectra. Remarkably, \Hy affects differently the FM and AFM samples, with opposing trends in the spectral weight change of the conduction band. Behind the observed behavior lies a nontrivial interaction mechanism which calls for future investigation. Overall, our work highlights the novelty and rich phenomenology of hydrogen incorporation as a controlled and effective way to manipulate the electronic properties of quantum systems.

\begin{acknowledgments}

 We acknowledge financial support  and of the Region Île de France via (FemtoARPES2.0 project N. EX079194). We thank synchrotron SOLEIL for providing beamtime under proposals 20250616 and 20252090.

\end{acknowledgments}

\newpage


\bibliographystyle{apsrev4-2}
\bibliography{apssamp}

@article{Li2019rev,
   author = {Jiaheng Li and Yang Li and Shiqiao Du and Zun Wang and Bing Lin Gu and Shou Cheng Zhang and Ke He and Wenhui Duan and Yong Xu},
   doi = {10.1126/sciadv.aaw5685},
   issn = {23752548},
   issue = {6},
   journal = {Science Advances},
   pages = {1-7},
   pmid = {31214654},
   title = {Intrinsic magnetic topological insulators in van der Waals layered MnBi2Te4-family materials},
   volume = {5},
   year = {2019}
}

@article{Chen2019,
   author = {Y. J. Chen and L. X. Xu and J. H. Li and Y. W. Li and H. Y. Wang and C. F. Zhang and H. Li and Y. Wu and A. J. Liang and C. Chen and S. W. Jung and C. Cacho and Y. H. Mao and S. Liu and M. X. Wang and Y. F. Guo and Y. Xu and Z. K. Liu and L. X. Yang and Y. L. Chen},
   doi = {10.1103/PhysRevX.9.041040},
   issn = {21603308},
   issue = {4},
   journal = {Physical Review X},
   pages = {1-9},
   title = {Topological Electronic Structure and Its Temperature Evolution in Antiferromagnetic Topological Insulator MnBi2Te4},
   volume = {91},
   year = {2019}
}

@article{Chen2019_2,
   author = {Bo Chen and Fucong Fei and Dongqin Zhang and Bo Zhang and Wanling Liu and Shuai Zhang and Pengdong Wang and Boyuan Wei and Yong Zhang and Zewen Zuo and Jingwen Guo and Qianqian Liu and Zilu Wang and Xuchuan Wu and Junyu Zong and Xuedong Xie and Wang Chen and Zhe Sun and Shancai Wang and Yi Zhang and Minhao Zhang and Xuefeng Wang and Fengqi Song and Haijun Zhang and Dawei Shen and Baigeng Wang},
   doi = {10.1038/s41467-019-12485-y},
   issn = {20411723},
   issue = {1},
   journal = {Nature Communications},
   pages ={},
   month = {12},
   pmid = {31578337},
   publisher = {Nature Publishing Group},
   title = {Intrinsic magnetic topological insulator phases in the Sb doped MnBi2Te4 bulks and thin flakes},
   volume = {10},
   year = {2019}
}

@article{Li2019,
   author = {Hang Li and Shun Ye Gao and Shao Feng Duan and Yuan Feng Xu and Ke Jia Zhu and Shang Jie Tian and Jia Cheng Gao and Wen Hui Fan and Zhi Cheng Rao and Jie Rui Huang and Jia Jun Li and Da Yu Yan and Zheng Tai Liu and Wan Ling Liu and Yao Bo Huang and Yu Liang Li and Yi Liu and Guo Bin Zhang and Peng Zhang and Takeshi Kondo and Shik Shin and He Chang Lei and You Guo Shi and Wen Tao Zhang and Hong Ming Weng and Tian Qian and Hong Ding},
   doi = {10.1103/PhysRevX.9.041039},
   issn = {21603308},
   issue = {4},
   journal = {Physical Review X},
   pages ={},
   month = {11},
   publisher = {American Physical Society},
   title = {Dirac Surface States in Intrinsic Magnetic Topological Insulators EuSn2As2 and MnBi2nTe3n+1},
   volume = {91},
   year = {2019}
}

@article{Swatek2020,
   author = {Przemyslaw Swatek and Yun Wu and Lin Lin Wang and Kyungchan Lee and Benjamin Schrunk and Jiaqiang Yan and Adam Kaminski},
   doi = {10.1103/PhysRevB.101.161109},
   issn = {24699969},
   issue = {16},
   journal = {Physical Review B},
   pages = {},
   month = {4},
   publisher = {American Physical Society},
   title = {Gapless Dirac surface states in the antiferromagnetic topological insulator MnBi2Te4},
   volume = {101},
   year = {2020}
}

@article{Zhang2011,
   author = {Jinsong Zhang and Cui Zu Chang and Zuocheng Zhang and Jing Wen and Xiao Feng and Kang Li and Minhao Liu and Ke He and Lili Wang and Xi Chen and Qi Kun Xue and Xucun Ma and Yayu Wang},
   doi = {10.1038/ncomms1588},
   issn = {20411723},
   issue = {1},
   journal = {Nature Communications},
   pages={},
   pmid = {22146393},
   title = {Band structure engineering in (Bi1-xSbx) 2Te3 ternary topological insulators},
   volume = {2},
   year = {2011}
}

@article{Kiyoi2026,
   author = {Akira Kiyoi and Yusuke Nishiya and Yuichiro Matsushita and Takahide Umeda},
   doi = {10.1038/s43246-025-00955-4},
   issn = {26624443},
   issue = {1},
   journal = {Communications Materials},
   pages = {},
   month = {12},
   publisher = {Springer Nature},
   title = {Advancing N-type doping in semiconductors through hydrogen-defect interactions},
   volume = {7},
   year = {2026}
}

@article{Myers1992,
  title = {Hydrogen interactions with defects in crystalline solids},
  author = {Myers, S. M. and Baskes, M. I. and Birnbaum, H. K. and Corbett, J. W. and DeLeo, G. G. and Estreicher, S. K. and Haller, E. E. and Jena, P. and Johnson, N. M. and Kirchheim, R. and Pearton, S. J. and Stavola, M. J.},
  journal = {Rev. Mod. Phys.},
  volume = {64},
  issue = {2},
  pages = {559--617},
  numpages = {0},
  year = {1992},
  month = {Apr},
  publisher = {American Physical Society},
  doi = {10.1103/RevModPhys.64.559},
  url = {https://link.aps.org/doi/10.1103/RevModPhys.64.559}
}

@article{Pearton1987,
  author  = {Pearton, S. J. and Corbett, J. W. and Shi, T. S.},
  title   = {Hydrogen in crystalline semiconductors},
  journal = {Applied Physics A},
  year    = {1987},
  volume  = {43},
  number  = {3},
  pages   = {153--195},
  doi     = {10.1007/BF00615975},
  url     = {https://doi.org/10.1007/BF00615975}
}

@article{Elias2009,
   author = {D C Elias and R R Nair and T M G Mohiuddin and S V Morozov and P Blake and M P Halsall and A C Ferrari and D W Boukhvalov and M I Katsnelson and A K Geim and K S Novoselov},
   doi = {10.1126/science.l092905},
   isbn = {6300100502},
   issue = {5914},
   journal = {New Series},
   pages = {610-613},
   title = {Control of Graphene's Properties by Reversible Hydrogenation: Evidence for Graphane},
   volume = {323},
   year = {2009}
}

@article{Shibuya2016,
   author = {Keisuke Shibuya and Akihito Sawa},
   doi = {10.1002/aelm.201500131},
   issn = {2199160X},
   issue = {2},
   journal = {Advanced Electronic Materials},
   pages = {},
   month = {2},
   publisher = {Blackwell Publishing Ltd},
   title = {Modulation of Metal–Insulator Transition in VO2 by Electrolyte Gating-Induced Protonation},
   volume = {2},
   year = {2016}
}

@article{Wang2019,
   author = {Meng Wang and Xuelei Sui and Yujia Wang and Yung Hsiang Juan and Yingjie Lyu and Huining Peng and Tongtong Huang and Shengchun Shen and Chenguang Guo and Jianbing Zhang and Zhuolu Li and Hao Bo Li and Nianpeng Lu and Alpha T. N'Diaye and Elke Arenholz and Shuyun Zhou and Qing He and Ying Hao Chu and Wenhui Duan and Pu Yu},
   doi = {10.1002/adma.201900458},
   issn = {15214095},
   issue = {16},
   journal = {Advanced Materials},
   pages = {},
   month = {4},
   pmid = {30811706},
   publisher = {Wiley-VCH Verlag},
   title = {Manipulate the Electronic and Magnetic States in NiCo2O4 Films through Electric-Field-Induced Protonation at Elevated Temperature},
   volume = {31},
   year = {2019}
}

@article{Li2023,
   author = {Zhuolu Li and Yingjie Lyu and Zhao Ran and Yujia Wang and Yang Zhang and Nianpeng Lu and Meng Wang and Michel Sassi and Thai Duy Ha and Alpha T. N'Diaye and Padraic Shafer and Carolyn Pearce and Kevin Rosso and Elke Arenholz and Jenh Yih Juang and Qing He and Ying Hao Chu and Weidong Luo and Pu Yu},
   doi = {10.1002/adfm.202212298},
   issn = {16163028},
   issue = {17},
   journal = {Advanced Functional Materials},
   pages = {},
   month = {4},
   publisher = {John Wiley and Sons Inc},
   title = {Enhanced Magnetization in CoFe2O4 Through Hydrogen Doping},
   volume = {33},
   year = {2023}
}

@article{Hu2024,
   author = {Xinmeng Hu and Xinyi He and Zhilin Guo and Toshio Kamiya and Jiazhen Wu},
   doi = {10.1021/acsnano.3c09064},
   issn = {1936086X},
   issue = {1},
   journal = {ACS Nano},
   month = {1},
   pages = {738-749},
   pmid = {38127649},
   publisher = {American Chemical Society},
   title = {Antisite-Defects Control of Magnetic Properties in MnSb2Te4},
   volume = {18},
   year = {2024}
}

@article{Yan2021,
   author = {D. Y. Yan and M. Yang and P. B. Song and Y. T. Song and C. X. Wang and C. J. Yi and Y. G. Shi},
   doi = {10.1103/PhysRevB.103.224412},
   issn = {24699969},
   issue = {22},
   journal = {Physical Review B},
   month = {6},
   pages={},
   publisher = {American Physical Society},
   title = {Site mixing induced ferrimagnetism and anomalous transport properties of the Weyl semimetal candidate MnSb2Te4},
   volume = {103},
   year = {2021}
}

@article{Deng2022,
   author = {Haiming Deng and Lukas Zhao and Kyungwha Park and Jiaqiang Yan and Kamil Sobczak and Ayesha Lakra and Entela Buzi and Lia Krusin-Elbaum},
   doi = {10.1038/s41467-022-29957-3},
   issn = {20411723},
   issue = {1},
   journal = {Nature Communications},
   pages = {1-8},
   pmid = {35484140},
   publisher = {Springer US},
   title = {Topological surface currents accessed through reversible hydrogenation of the three-dimensional bulk},
   volume = {13},
   year = {2022}
}

@article{Tamanna2024,
   author = {Afrin N. Tamanna and Ayesha Lakra and Xiaxin Ding and Entela Buzi and Kyungwha Park and Kamil Sobczak and Haiming Deng and Gargee Sharma and Sumanta Tewari and Lia Krusin-Elbaum},
   doi = {10.1038/s41467-024-53319-w},
   issn = {20411723},
   issue = {1},
   journal = {Nature Communications},
   month = {12},
   pmid = {39537597},
   pages = {},
   publisher = {Nature Research},
   title = {Transport chirality generated by a tunable tilt of Weyl nodes in a van der Waals topological magnet},
   volume = {15},
   year = {2024}
}

@article{Avila2023,
   author = {J. Avila and S. Lorcy and P. Dudin},
   doi = {10.1016/j.elspec.2023.147362},
   issn = {03682048},
   journal = {Journal of Electron Spectroscopy and Related Phenomena},
   month = {7},
   pages = {},
   publisher = {Elsevier B.V.},
   title = {ANTARES: Space-resolved electronic structure},
   volume = {266},
   year = {2023}
}

@article{Faure2012,
   author = {J. Faure and J. Mauchain and E. Papalazarou and W. Yan and J. Pinon and M. Marsi and L. Perfetti},
   doi = {10.1063/1.3700190},
   issn = {00346748},
   issue = {4},
   journal = {Review of Scientific Instruments},
   month = {4},
   pages = {},
   publisher = {Review of Scientific Instruments},
   title = {Full characterization and optimization of a femtosecond ultraviolet laser source for time and angle-resolved photoemission on solid surfaces},
   volume = {83},
   year = {2012}
}

@article{Huan2021,
   author = {Shuchun Huan and Shihao Zhang and Zhicheng Jiang and Hao Su and Hongyuan Wang and Xin Zhang and Yichen Yang and Zhengtai Liu and Xia Wang and Na Yu and Zhiqiang Zou and Dawei Shen and Jianpeng Liu and Yanfeng Guo},
   doi = {10.1103/PhysRevLett.126.246601},
   issn = {10797114},
   issue = {24},
   journal = {Physical Review Letters},
   month = {6},
   pmid = {34213928},
   pages={},
   publisher = {American Physical Society},
   title = {Multiple Magnetic Topological Phases in Bulk van der Waals Crystal MnSb4Te7},
   volume = {126},
   year = {2021}
}

@article{Xi2025,
   author = {Ming Xi and Yuchong Zhang and Wenju Zhou and Famin Chen and Donghan Jia and Huiyang Gou and Tian Qian and Hechang Lei},
   doi = {10.1021/acs.jpclett.5c02250},
   issn = {19487185},
   issue = {36},
   journal = {The journal of physical chemistry letters},
   month = {9},
   pages = {9507-9516},
   pmid = {40902191},
   title = {Relationship among Structure, Disorder, Magnetism, and Band Topology in the MnSb2Te4·(Sb2Te3)n Family},
   volume = {16},
   year = {2025}
}

@article{Xiong2023,
   author = {Jie Xiong and Yin Hui Peng and Jia Yi Lin and Yu Jie Cen and Xiao Bao Yang and Yu Jun Zhao},
   doi = {10.3390/ma16155496},
   issn = {19961944},
   issue = {15},
   journal = {Materials},
   month = {8},
   pages = {},
   publisher = {Multidisciplinary Digital Publishing Institute (MDPI)},
   title = {High Concentration Intrinsic Defects in MnSb2Te4},
   volume = {16},
   year = {2023}
}

@article{Zhang2023,
   author = {Jiuxiang Zhang and Thibault Daniel Pierre Sohier and Michele Casula and Zhesheng Chen and Jonathan Caillaux and Evangelos Papalazarou and Luca Perfetti and Luca Petaccia and Azzedine Bendounan and Amina Taleb-Ibrahimi and David Santos-Cottin and Yannick Klein and Andrea Gauzzi and Marino Marsi},
   doi = {10.1021/acs.nanolett.2c04701},
   issn = {15306992},
   issue = {5},
   journal = {Nano Letters},
   month = {3},
   pages = {1830-1835},
   pmid = {36651800},
   publisher = {American Chemical Society},
   title = {Manipulating Dirac States in BaNiS2 by Surface Charge Doping},
   volume = {23},
   year = {2023}
}

@article{Pinto2014,
   author = {Hugo Pinto and Alexander Markevich},
   doi = {10.3762/bjnano.5.195},
   issn = {21904286},
   issue = {1},
   journal = {Beilstein Journal of Nanotechnology},
   volume = {5},
   issue = {},
   pages = {1842-1848},
   publisher = {Beilstein-Institut Zur Forderung der Chemischen Wissenschaften},
   title = {Electronic and electrochemical doping of graphene by surface adsorbates},
   year = {2014}
}

@article{Iwanowski2010,
   author = {R. J. Iwanowski and M. H. Heinonen and B. Witkowska},
   doi = {10.1016/j.jallcom.2009.10.223},
   issn = {09258388},
   issue = {1-2},
   journal = {Journal of Alloys and Compounds},
   month = {2},
   pages = {13-17},
   title = {X-ray photoelectron study of NiAs-type MnTe},
   volume = {491},
   year = {2010}
}

@article{Li2022,
   author = {Hao Li and Yaoxin Li and Yukun Lian and Weiwei Xie and Ling Chen and Jinsong Zhang and Yang Wu and Shoushan Fan},
   doi = {10.1007/s40843-021-1738-9},
   issn = {21994501},
   issue = {2},
   journal = {Science China Materials},
   month = {2},
   pages = {477-485},
   publisher = {Science Press (China)},
   title = {Glassy magnetic ground state in layered compound MnSb2Te4},
   volume = {65},
   year = {2022}
}

@article{Liu2021,
   author = {Yaohua Liu and Lin Lin Wang and Qiang Zheng and Zengle Huang and Xiaoping Wang and Miaofang Chi and Yan Wu and Bryan C. Chakoumakos and Michael A. McGuire and Brian C. Sales and Weida Wu and Jiaqiang Yan},
   doi = {10.1103/PhysRevX.11.021033},
   issn = {21603308},
   issue = {2},
   journal = {Physical Review X},
   note = {},
   pages = {21033},
   publisher = {American Physical Society},
   title = {Site Mixing for Engineering Magnetic Topological Insulators},
   volume = {11},
   url = {https://doi.org/10.1103/PhysRevX.11.021033},
   year = {2021}
}

@article{Yan2019,
   author = {J. Q. Yan and S. Okamoto and M. A. McGuire and A. F. May and R. J. McQueeney and B. C. Sales},
   doi = {10.1103/PhysRevB.100.104409},
   issn = {24699969},
   issue = {10},
   journal = {Physical Review B},
   month = {9},
   pages = {},
   publisher = {American Physical Society},
   title = {Evolution of structural, magnetic, and transport properties in MnBi2-xSbxTe4},
   volume = {100},
   year = {2019}
}

@article{Eremeev2017, 
   author = {S. V. Eremeev and M. M. Otrokov and E. V. Chulkov},
   doi = {10.1016/j.jallcom.2017.03.121},
   issn = {09258388},
   journal = {Journal of Alloys and Compounds},
   pages = {172-178},
   publisher = {Elsevier Ltd},
   title = {Competing rhombohedral and monoclinic crystal structures in MnPn2Ch4 compounds: An ab-initio study},
   volume = {709},
   year = {2017}
}

@article{Eremeev2021,
   author = {S. V. Eremeev and I. P. Rusinov and Yu M. Koroteev and A. Yu Vyazovskaya and M. Hoffmann and P. M. Echenique and A. Ernst and M. M. Otrokov and E. V. Chulkov},
   doi = {10.1021/acs.jpclett.1c00875},
   issn = {19487185},
   issue = {17},
   journal = {Journal of Physical Chemistry Letters},
   month = {5},
   pages = {4268-4277},
   pmid = {33908787},
   publisher = {American Chemical Society},
   title = {Topological magnetic materials of the (MnSb2Te4)·(Sb2Te3)nvan der waals compounds family},
   volume = {12},
   year = {2021}
}

@article{Zhou2020,
   author = {Liqin Zhou and Zhiyun Tan and Dayu Yan and Zhong Fang and Youguo Shi and Hongming Weng},
   doi = {10.1103/PhysRevB.102.085114},
   issn = {24699969},
   issue = {8},
   journal = {Physical Review B},
   note = {},
   pages = {85114},
   publisher = {American Physical Society},
   title = {Topological phase transition in the layered magnetic compound MnSb2Te4: Spin-orbit coupling and interlayer coupling dependence},
   volume = {102},
   url = {https://doi.org/10.1103/PhysRevB.102.085114},
   year = {2020}
}

@article{Li2021,
   author = {Ping Li and Jiangying Yu and Ying Wang and Weidong Luo},
   doi = {10.1103/PhysRevB.103.155118},
   issn = {24699969},
   issue = {15},
   journal = {Physical Review B},
   note = {},
   pages = {155118},
   publisher = {American Physical Society},
   title = {Electronic structure and topological phases of the magnetic layered materials MnBi2 Te4, MnBi2 Se4, and MnSb2 Te4 ELECTRONIC STRUCTURE and TOPOLOGICAL PHASES ... LI, YU, WANG, and LUO},
   volume = {103},
   url = {https://doi.org/10.1103/PhysRevB.103.155118},
   year = {2021}
}

@article{Chen2020,
   author = {Li Chen and Dongchao Wang and Changmin Shi and Chuan Jiang and Hongmei Liu and Guangliang Cui and Xiaoming Zhang and Xiaolong Li},
   doi = {10.1007/s10853-020-05005-7},
   issn = {15734803},
   issue = {29},
   journal = {Journal of Materials Science},
   note = {},
   pages = {14292-14300},
   title = {Electronic structure and magnetism of MnSb2Te4},
   volume = {55},
   year = {2020}
}

@article{Murakami2019,
   author = {Taito Murakami and Yusuke Nambu and Takashi Koretsune and Gu Xiangyu and Takafumi Yamamoto and Craig M. Brown and Hiroshi Kageyama},
   doi = {10.1103/PhysRevB.100.195103},
   issn = {24699969},
   issue = {19},
   journal = {Physical Review B},
   note = {},
   pages = {195103},
   publisher = {American Physical Society},
   title = {Realization of interlayer ferromagnetic interaction in MnS b2 T e4 toward the magnetic Weyl semimetal state},
   volume = {100},
   url = {https://doi.org/10.1103/PhysRevB.100.195103},
   year = {2019}
}

@article{Xi2022,
   author = {Ming Xi and Famin Chen and Chunsheng Gong and Shangjie Tian and Qiangwei Yin and Tian Qian and Hechang Lei},
   issue = {47},
   pages={},
   journal = {Journal of Physical Chemistry Letters},
   pages = {10897-10904},
   title = {{Relationship between Antisite Defects, Magnetism, and Band Topology in MnSb2Te4 Crystals with TC≈ 40 K}},
   volume = {13},
   year = {2022},
   doi = {10.1021/acs.jpclett.2c02775}
}

@Article{Zhao2016,
author={Zhao, Lukas
and Konczykowski, Marcin
and Deng, Haiming
and Korzhovska, Inna
and Begliarbekov, Milan
and Chen, Zhiyi
and Papalazarou, Evangelos
and Marsi, Marino
and Perfetti, Luca
and Hruban, Andrzej
and Wo{\l}o{\'{s}}, Agnieszka
and Krusin-Elbaum, Lia},
title={Stable topological insulators achieved using high energy electron beams},
journal={Nature Communications},
year={2016},
month={Mar},
day={10},
volume={7},
number={1},
pages={10957},
issn={2041-1723},
doi={10.1038/ncomms10957},
url={https://doi.org/10.1038/ncomms10957}
}

\end{document}